\documentclass[]{spie}  

\usepackage[utf8]{inputenc}
\usepackage{amsmath,amsfonts,amssymb}
\usepackage{gensymb}
\usepackage{graphicx}
\usepackage[colorlinks=true,allcolors=blue]{hyperref}
\newcommand{\fdg}{\mbox{\ensuremath{.\!\!^{\degree}}}}%
\newcommand{\farcs}{\mbox{\ensuremath{.\!\!^{\prime\prime}}}}%

\title{TAUKAM: A 6k$\,\times\,$6k
prime-focus camera for the Tautenburg Schmidt Telescope}

\author[a]{Bringfried Stecklum}
\author[a]{Jochen Eislöffel}
\author[a]{Sylvio Klose}
\author[a]{Uwe Laux}
\author[a]{Tom Löwinger}
\author[a]{Helmut Meusinger}
\author[a]{Michael Pluto}
\author[a]{Johannes Winkler}
\author[b]{Frank Dionies}
\affil[a]{Thüringer Landessternwarte Tautenburg, Sternwarte 5, 07778 Tautenburg, Germany}
\affil[b]{Leibniz Institut für Astrophysik, An der Sternwarte 16, 14482 Potsdam, Germany}
\authorinfo{Further author information: (Send correspondence to B.S.)\\B.S.: E-mail: stecklum@tls-tautenburg.de, Phone: +49 36427 863 54}

\begin{document} 
\maketitle

\begin{abstract}

TAUKAM stands for "TAUtenburg KAMera", which will become the new prime-focus imager for the Tautenburg Schmidt telescope.
It employs an e2v 6k\,$\times$\,6k CCD and is under manufacture by Spectral Instruments Inc. We describe the design of the instrument and the auxiliary components, its specifications as well as the concept for integrating the device into the  telescope infrastructure. First light is foreseen in 2017. TAUKAM will boost the observational capabilities of the telescope for what concerns optical wide-field surveys.
\end{abstract}


\section{INTRODUCTION}
\label{sec:intro}  

The 2-m telescope of the Karl Schwarzschild Observatory, Tautenburg 
(IAU station code 033) -- 
which became the Thüringer Landessternwarte (TLS) after the German re-unification --
was built by Carl Zeiss Jena and went into operation in 1960 \cite{1961Obs....81...91V}.
It is a versatile device which offers three optical configurations (Coude, Nasmyth, Schmidt).  
The Schmidt mode utilizes the 1.34-m correction plate. In this mode the telescope still represents the largest {\em{imaging}} Schmidt system with a vignette-free field of view (FOV) of 3\fdg3$\,\times\,$3\fdg3, and is regularly used during dark time for wide-field imaging (typically $\pm$1 week around New Moon). The exchange of the photographic plate holder by a prime focus CCD camera in 1996 led to a fractional coverage of the FOV only. 
Replacing the outdated 2k\,$\times$\,2k CCD camera by a more powerful one was urgently needed to foster our research programs which are tailored to the local observing conditions ($\sim$1000$\pm$200 observing hours per year at a median seeing of $\sim$2$^{\prime\prime}$). These comprise, e.g., variability studies of young stars and quasars, target of opportunity observations, and follow-up of near-Earth objects (NEOs). In particular, the slow read-out electronics of the old camera leads to substantial overheads. Moreover, its pixel scale of 1\farcs24 implies undersampling of the point spread function (PSF) in case of good seeing conditions. When the Thuringian Ministry for Education, Science, and Culture issued a call for proposals in 2014 within the framework of their infrastructure development program, we submitted an application for a new instrument, named TAUKAM. For reasons becoming more clear in the following, the 1110S CCD camera from Spectral Instruments Inc. (SI) was used as the reference case in our proposal. The latter was approved which allowed us to submit a Europe-wide call for tender in 2015. 
The successful bid was offered by Photon Lines SAS which is the French distributor of products from SI. The contract, signed in September 2015, includes the delivery of a 1110S CCD camera equipped with an e2v 6k$\,\times\,$6k detector in the beginning of 2017. It is foreseen to start regular observations at the end of that year.
In the following sections we describe the technical and scientific specifications of TAUKAM, the design of the dewar including the entry window as well as the concepts to integrate the instrument into the telescope infrastructure. While various technical components are fixed already, others are still being developed or under consideration. Thus, the present publication represents rather a summary of the implementation of TAUKAM than a final report. 


\section{Telescope optics and tube}
\begin{figure}[ht]
\begin{center}
\includegraphics[trim=50mm 90mm 0mm 90mm, clip, height=4.5cm]{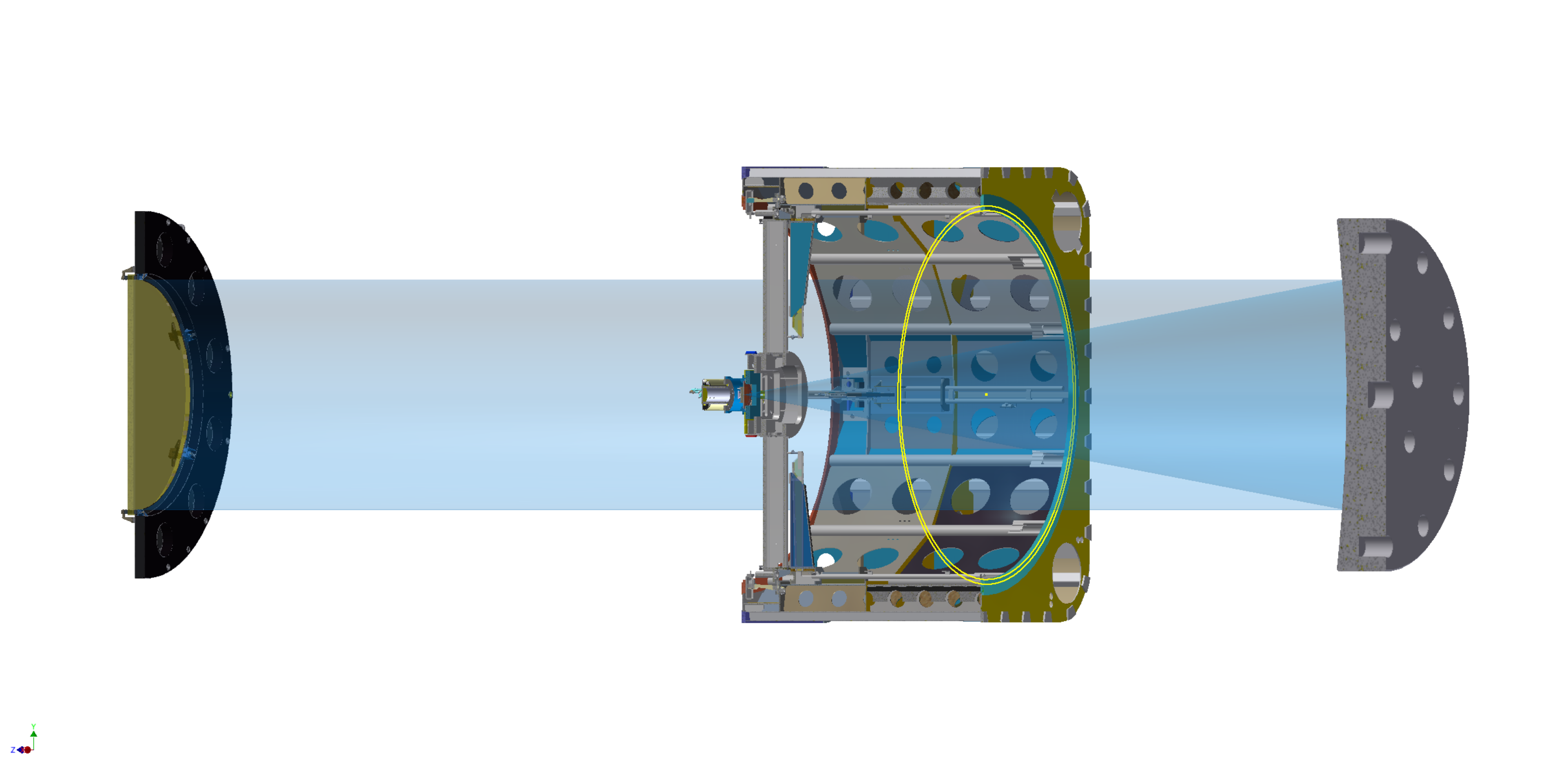}
\end{center}
\caption{Cut of the Schmidt optics with the 1110S camera mounted at the focal plane flange. Light is passing the Schmidt plate (left, light brown), gets reflected at the spherical mirror (right), and is focused onto the CCD. Only the tube section close to the focal plane is shown.
\label{fig:optics} }
\end{figure}

The Tautenburg telescope represents a classical Schmidt system with a spherical 2-m main mirror and a \mbox{1.34-m} correction lens (focal ratio f/3, see Fig.\,\ref{fig:optics}). Apart from the Schmidt mode it also offers Nasmyth and Coude configurations where the latter is used during bright time for high-resolution spectroscopy. Being a classical Schmidt telescope it features a closed tube of square cross section. The telescope guiding is realized with a separate 0.3-m refractor housed in the tube which hosts an intensified video camera. The closed tube has to be taken into account for what concerns the heat dissipation of TAUKAM. Unlike the former CCD cameras its read-out electronics will be integrated in the dewar, i.e. located close to the focal plane. The effect of the dissipation of the predicted power of $\sim$35\,W on both image quality and tube temperature was tested experimentally using a device of similar power consumption. While no adverse affects were noticed, a preventive countermeasure is to allow for air circulation by opening the lateral tube sliders which were formerly used to insert the photographic plate holder. 

\begin{table}[b]
\caption{Detector specifications}
\label{ccd:tab1}
\vspace*{-2mm}
\begin{center}
\begin{tabular}{ll}
\hline\hline
\noalign{\vskip 1mm}
CCD format & 6144\,(H)\,$\times$\,6160\,(V) pixel\\
Image area & 92.2\,mm\,$\times$\,92.4\,mm\\
Pixel size and scale & 15\,$\mu$m, 0\farcs771 \\
FoV (unvignetted) &  1.73\,${^\square}\degree$\\
Read speeds & $\sim$100, $\sim$433, and 752\,kHz\\
Read-out times (using 4 ports) & 94, 21, and 12.5\,s\\
Read-out noise at 100, 752\,kHz & goal $<$2.5, 4.1\,e$^-$\\
Full-well capacity & 150,000\,e$^-$\\
Operating temperature & $-110$\dots$ -100\,\degree$C\\
Dark current & 0.0008 $\rm e^- pix^- s^-$\\
Flatness & $<$40\,$\mu$m (peak to valley) \\
\hline
\end{tabular}
\end{center}
\end{table} 

\section{CCD detector}
The detector of TAUKAM will be a  6144\,(H)\,$\times$\,6160\,(V)  back-illuminated scientific CCD sensor (CCD-231-C6-1-G11) produced by e2v, Chelmsford (UK). It is of deep-depletion silicon type, features four outputs with 16 bit analog-to-digital (AD) conversion, and is operated in non-inverted mode. The pixel size of 15\,$\mu$m corresponds to  0\farcs771 which ensures proper image sampling for the prevailing seeing conditions. The silicon carbide package provides a compact footprint with guaranteed flatness at cryogenic temperatures. The selected astro-multi-2 AR coating provides more than 50\,\% quantum efficiency over a wavelength range from 350 to 920 nanometer. Using this detector TAUKAM will provide a FOV of 1.73\,${^\square}\degree$,  a more than fourfold increase compared to its predecessor. The full-frame read-out time amounts to about half of that of the present device and may be even much shorter, which implies a large overhead reduction. Table~\ref{ccd:tab1} summarizes major specifications of the detector.

\section{CCD Dewar}

\subsection{Dewar window and design}
\begin{figure}[ht]
\begin{center}
\begin{tabular}{cc} 
\includegraphics[trim=125mm 0mm 0mm 0mm, clip, angle=90, height=5cm, width=8.25cm]{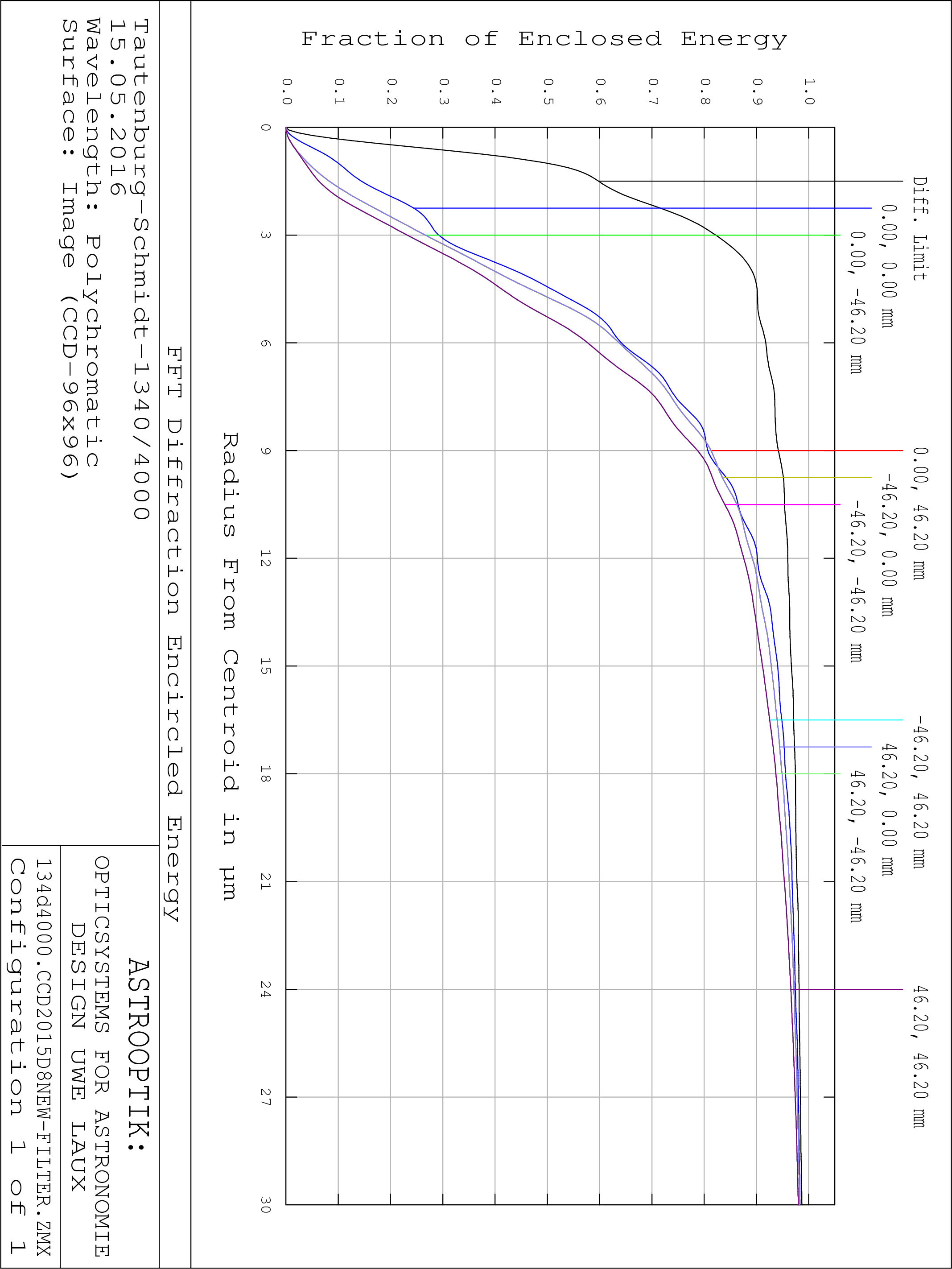}
\includegraphics[height=5cm, width=8.5cm]{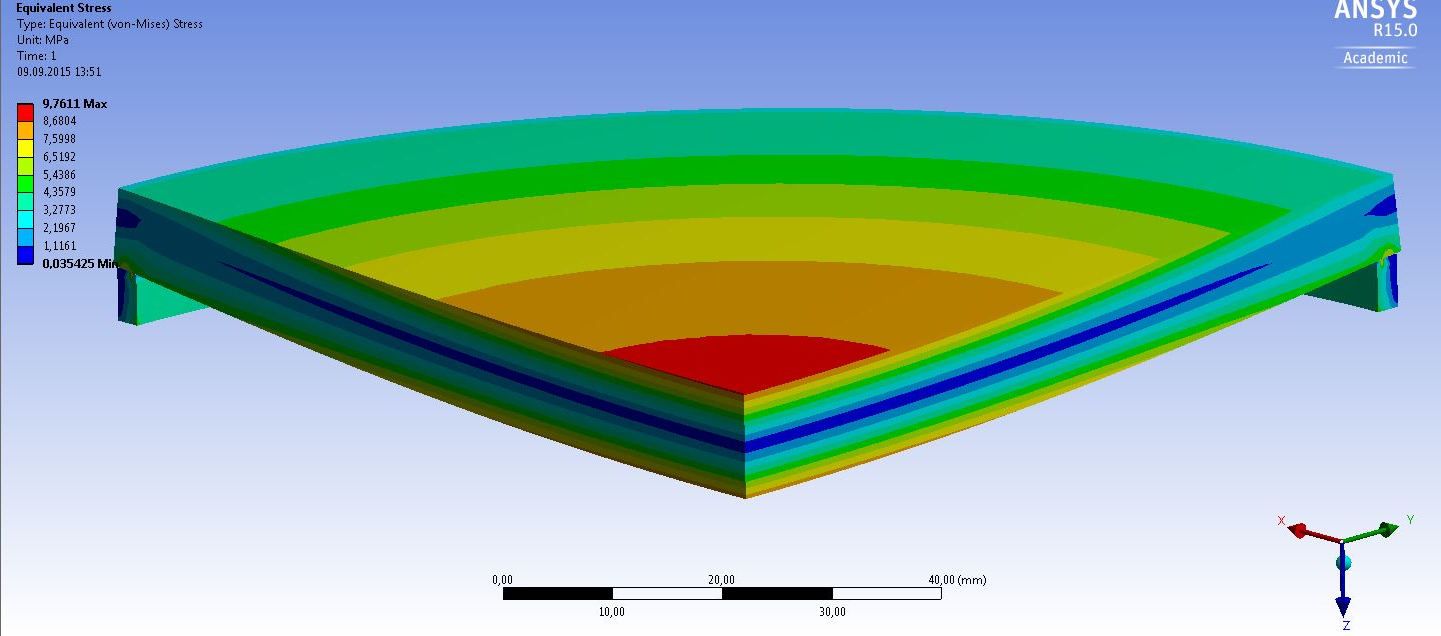}
\end{tabular}
\end{center}
\caption{Left: Encircled energy for center, edge mid-points, and corner positions of the field compared to the diffraction limit for polychromatic light. Right: Tension distribution across a quarter of the window.
\label{fig:window} }
\end{figure}

The dewar window represents a plano-convex lens which will correct the field 
curvature of the Schmidt focal plane to match the flat CCD surface. The optical design was performed using Zemax\footnote{Zemax is a trademark of Zemax LLC} and own software \cite{2002astr.book.....L}. The lens diameter amounts to 166\,mm and its curvature radius is 1349.2\,mm. Fig.\,\ref{fig:window} (left) shows the beam quality provided by the field lens for the center, edge mid-points, and corners of the detector. The mechanical stress test of the lens design was performed using a finite element method (FEM) analysis (Fig.\,\ref{fig:window} right). The peak tension of $\sim$10\,MPa due to the outside atmospheric pressure is much lower than the critical value of 54\,MPa for a substrate made from Corning 7980 fused silica. The lens was produced by Korth Kristalle GmbH, Altenholz (Germany), and shipped to the camera manufacturer for integration in early 2016. 

\subsection{Dewar body and mounting flange}
\label{dew:body}
\begin{figure}[h]
\begin{center}
\begin{tabular}{cc} 
\includegraphics[trim=8mm 45mm 7mm 23mm, clip, height=5cm]{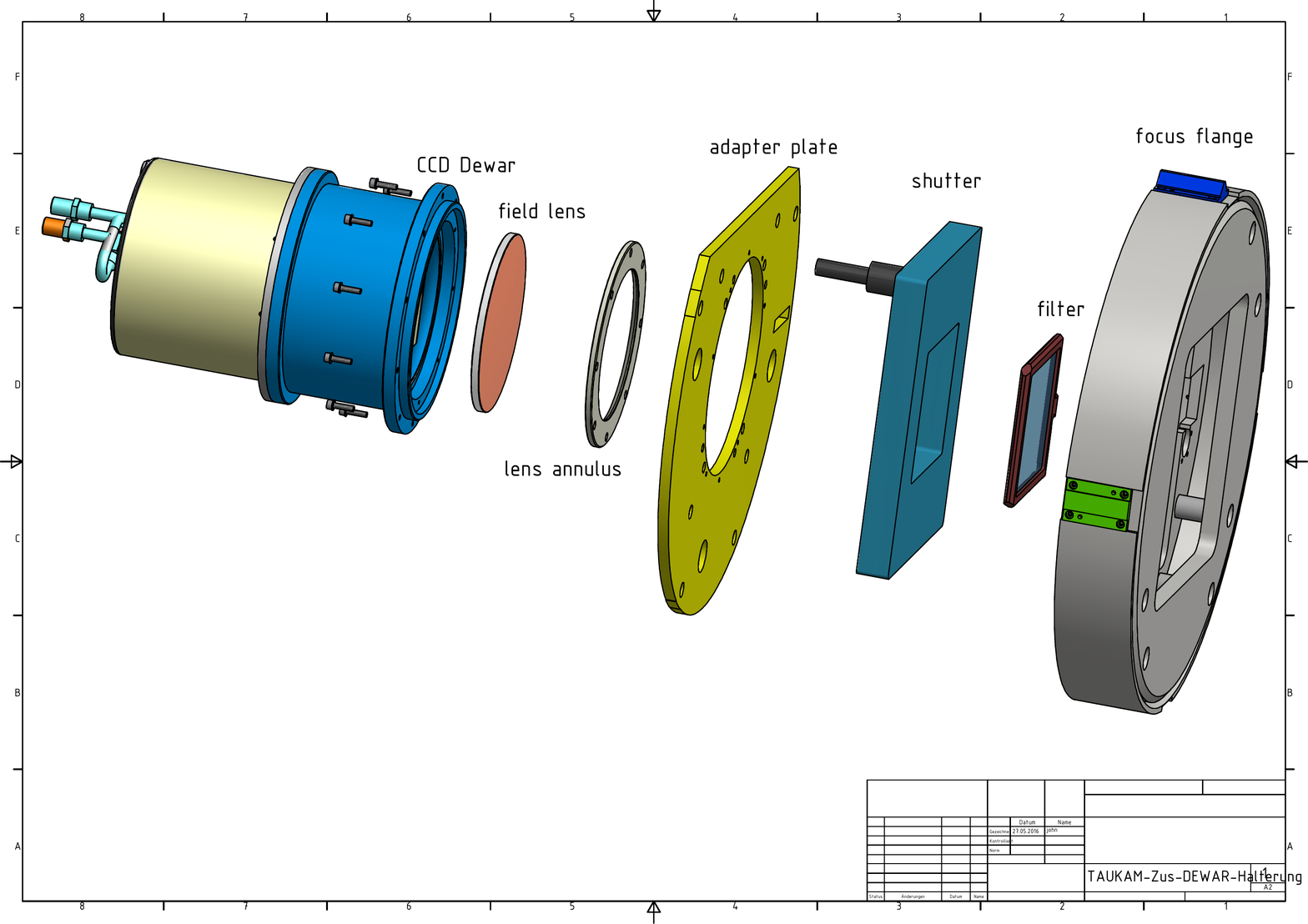}
\includegraphics[trim=100mm 0mm 100mm 0mm, clip,height=5cm]{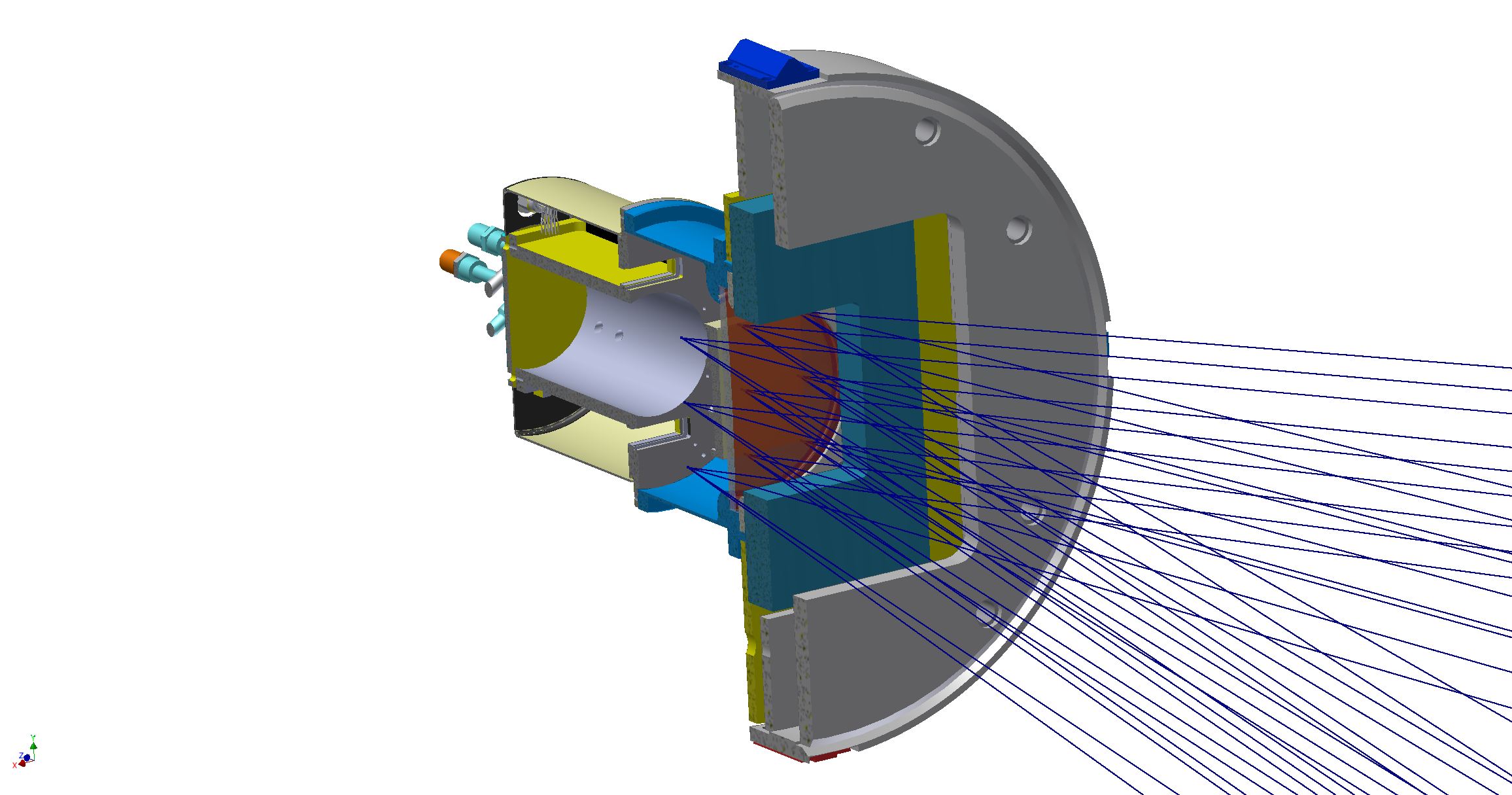}
\end{tabular}
\end{center}
\caption{Left: Exploded view of major camera components. The dewar head (blue) was designed to meet the adapter plate requirements. Right: Dewar with attached shutter mounted at the flange and incident rays.
\label{fig:dewar} }
\end{figure}
\noindent
While the filters for the old CCD camera are small enough to be housed in a filter wheel which fits behind the focus flange between the dewar and the adapter plate this is no longer possible with TAUKAM. Even populating two wheels with filters (and an empty placeholder) is no remedy since their diameter would still exceed that of the focus flange, leading to severe vignetting. Moreover, the presence of four bolts which serve for the movement of the focus flange prevents the use of filter wheels this big as well. Fig.\,\ref{fig:dewar} shows the major camera components along with a section view of the assembled state. For the new camera the filters have to be put in place from the opposite side of the focus flange (cf. Sect.\,\ref{sec:filt}). This will allow us to move the CCD slightly closer to the adapter plate, i.e. in a more advantageous focus range. Thanks to a re-design of the original dewar head, a reproducable and well-centered insertion of the dewar head into the adapter plate is guaranteed.

\section{Shutter}

The 1110S camera incorporates an internal flexible shutter drive to trigger an external shutter. The shutter will be located on the backside of the flange where the dewar is mounted (Fig.\,\ref{fig:dewar}). It needs to be compact as well to prevent beam vignetting. Its thickness is restricted to 30\,mm 
and according to the Schmidt optics the free opening has to be 115\,mm\,$\times$\,115\,mm.

The removal of an obsolete mechanical unit (used to widen photographic objective-prism spectra) provided extra space for the shutter. For what concerns commercially available devices, the dual double-blade compact shutter of Bonn Shutter UG, Bonn (Germany), meets these requirements. The mechanics is based on two carbon fiber or sandwich-type blades moving on a pair of linear ball bearings, driven by two stepper motors and toothed belts. The control electronics hardware consists of two identical micro-controller systems -- one for each shutter blade. The Bonn shutters are impact free, low acceleration (i.e. low power) devices. Negotiations with Bonn Shutter started in 2015 for what concerns the incorporation of this product. They have expertise with regard to driving their shutters with cameras from SI.

\section{Filters and filter unit}
\label{sec:filt}
The filter size amounts to 120\,mm\,$\times$\,120\,mm. With TAUKAM we will move away from Johnson-Cousins UBVRI glass filters in favor of the more advantageous Sloan Digital Sky Survey (SDSS) $u\,g\,r\,i\,z$ system. This filter set will be augmented by narrow-band filters (e.g., H$\alpha$, 
[S{\sc ii}]
) and a few others. A Johnson-Cousins $B$ filter will be added for the consistency of long-term photometric monitoring. Moreover, a broad-band $V$ filter ($VB$) will be utilized as well for the observation of NEOs where a large band-width is crucial to achieve a high signal-to-noise ratio (SNR). All filters will be based on multi-layer dielectric coatings and of same thickness, i.e. confocal. 

\begin{figure}[ht]
\begin{center}
\begin{tabular}{cc} 
\includegraphics[trim=110mm 100mm 60mm 100mm, clip, height=5cm]{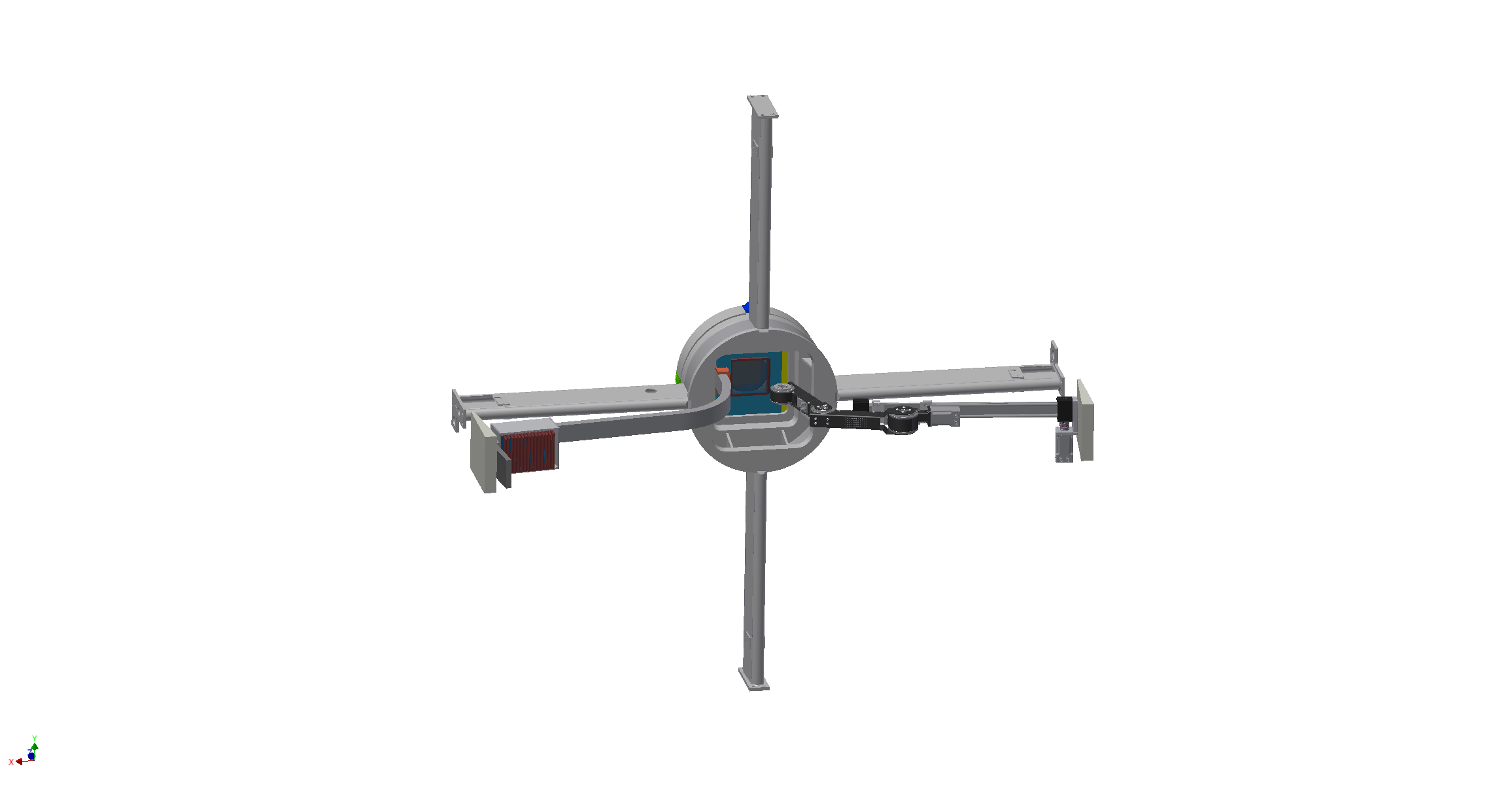}
\end{tabular}
\end{center}
\caption{Two concepts for the filter unit. Left: A rail mechanism carries a filter from the storage box located at the edge of the tube to the nominal position in front of the shutter. Right: The same task is performed using robotic lever arms in connection with torsion drives.
\label{fig:filter} }
\end{figure}

As outlined in Sect.\,\ref{dew:body} the use of wheels for housing the filters is not feasible. An alternative concept is a filter box located at the inner tube wall, i.e. out of the beam in Schmidt mode, from which filters will be grabbed and placed into position by a robotic unit (Fig.\,\ref{fig:filter}). The unit will hide behind one of the spider arms to prevent additional obscuration. It is intended that the filter exchange will be as fast as the shortest full-frame read-out time.

\section{Cryogenics}
Unlike the previous prime-focus CCD cameras which were cooled using liquid nitrogen (LN2), 
TAUKAM will be cooled by a cryo compressor which is part of the contract. It will be installed in the basement to avoid dome seeing due to the nominal power dissipation of 600\,W.  From there the cooling lines will run to the dewar inside the tube. For that reason three segments of cooling lines were ordered with a total length of 45.7\,m. The effective length change of the lines due to the telescope rotation will be canceled by cable twisters, similar to those for the GROND multi-channel imager \cite{2008PASP..120..405G} mounted at the Max Planck Society 2.2-m telescope on ESO/La Silla, Chile, or the 0.7-m Bigelow Schmidt telescope of the Catalina Sky Survey \cite{2015DPS....4730819C}.

The camera will be kept connected to the cooling unit when the telescope is operated in other optical configurations (Coude, Nasmyth). During these periods the dewar is foreseen to be stowed away in one corner of the tube.



\section{Computing and software}
\label{sec:misc}
The data are transferred in FITS format using a fiber optical cable to the image acquisition host computer via a proprietary PCIe card. For this purpose the accompanying software package {\em SI image II}
can be used. However, we will use the software development kit which is also provided by SI to integrate the camera control and data acquisition into our software environment.

\section{Scientific applications}
\label{sec:science}
In order to illustrate the impact of the TAUKAM imaging capabilities, their relevance for various science fields under investigation at TLS will be mentioned briefly.\\

\subsection{High-energy transients}
TLS joined the LIGO-Virgo Collaboration for follow-up observations of Gravitational Wave (GW) events in 2015, which includes a world-wide network of telescopes. The goal here is the search for optical transients following GW events that appear in the right time window. Since in the first years (2015+) the expected error boxes sizes are in the 100 to 1000 square degree range, telescopes with large field of views are required to find potential candidates. In addition, TLS has a long-standing activity in Gamma-ray burst (GRB) afterglow research, including observations with the Schmidt telescope. 
Error boxes here can be in the square degree range, too (mainly those coming from the Fermi satellite). As such, having a bigger field of view will represent a substantial advantage.  

\subsection{Long-term variability of quasars}
TLS carries out a unique, long-term variability study of $\sim$350 quasars with an epoch range of more than 50 years. This data base holds the potential to answer crucial questions concerning the time-scales of the quasar variability and its origin\cite{2010A&A...512A...1M}. While the collection of Schmidt plates represents the foundation for this project, the monitoring is continued with the prime-focus CCD camera for more than ten years now. However, the old CCD covers only 4\% of one monitoring field which leads to a sparse time sampling. Both the bigger FOV and the shorter read-out time of TAUKAM will improve the efficiency of the monitoring, and yield light curves of higher fidelity.

\subsection{Variability and rotation periods of young stars}
The removal of angular momentum during proto-stellar growth is one of the key issues of star formation. In order to gain insights into this process, we, for the first time, derived and studied rotation periods over a wide range of masses for objects of star clusters. These include so-called brown dwarfs which are{} ``failed stars" of very low mass, incapable of nuclear burning. Much to our surprise these tend to rotate faster than solar-like stars which indicates a less-efficient breaking during their formation\cite{2005A&A...429.1007S}. These ongoing investigations will profit from the capabilities of TAUKAM in particular for what concerns sky coverage and imaging duty cycle.

\subsection{Near-Earth objects }
TLS joined the NEO confirmation campaign coordinated by the Minor Planet Center (MPC), Cambridge (USA), on behalf of the IAU in 2010\cite{PDC2015}. While more than 3500 positions were reported to MPC since then, these are based on sub-frames which had to be taken to avoid the large read-out overhead of the old camera. This is particularly disadvantageous for the observations of newly-detected bodies (one-nighters) which often have large position errors. TAUKAM will allow us to generally employ the full-frame mode which implies an important increase of the observable NEO target sample, and improves chances for new asteroid discoveries by a large margin.

\acknowledgments 
 
We thank Eric Cristensen, University of Arizona, Catalina Sky Survey, for valuable advice. We are grateful for the cooperative exchange of information with Spectral Instruments Inc. and SAS Photon Lines.

\bibliography{main} 

\begin{thebibliography}{1}

\bibitem{1961Obs....81...91V}
{von Kluber}, H., ``{The new reflecting telescope at the Karl-Schwarzschild
  Observatory Tautenburg},'' {\em The Observatory}~{\bf 81},  91--94 (June
  1961).

\bibitem{2002astr.book.....L}
{Laux}, U.,  [{\em Astrooptik}{\nolinebreak\hspace{0.1em}]}, Verlag Sterne und
  Weltraum, Hüthig GmbH, second~ed. (2001).

\bibitem{2008PASP..120..405G}
{Greiner}, J., {Bornemann}, W., {Clemens}, C., {Deuter}, M., {Hasinger}, G.,
  {Honsberg}, M., {Huber}, H., {Huber}, S., {Krauss}, M., {Kr{\"u}hler}, T.,
  {K{\"u}pc{\"u} Yolda{\c s}}, A., {Mayer-Hasselwander}, H., {Mican}, B.,
  {Primak}, N., {Schrey}, F., {Steiner}, I., {Szokoly}, G., {Th{\"o}ne}, C.~C.,
  {Yolda{\c s}}, A., {Klose}, S., {Laux}, U., and {Winkler}, J.,
  ``{GROND\,{\mdash}\,a 7-Channel Imager},'' {\em \pasp}~{\bf 120},  405--424
  (Apr. 2008).

\bibitem{2015DPS....4730819C}
{Christensen}, E.~J., {Carson Fuls}, D., {Gibbs}, A.~R., {Grauer}, A.~D.,
  {Hill}, R.~E., {Johnson}, J.~A., {Kowalski}, R.~A., {Larson}, S.~M.,
  {Matheny}, R.~G., and {Shelly}, F.~C., ``{Status of The Catalina Sky
  Survey},'' in [{\em AAS/Division for Planetary Sciences Meeting
  Abstracts}{\nolinebreak\hspace{0.1em}]},  {\em AAS/Division for Planetary
  Sciences Meeting Abstracts} {\bf 47},  308.19 (Nov. 2015).

\bibitem{2010A&A...512A...1M}
{Meusinger}, H., {Henze}, M., {Birkle}, K., {Pietsch}, W., {Williams}, B.,
  {Hatzidimitriou}, D., {Nesci}, R., {Mandel}, H., {Ertel}, S., {Hinze}, A.,
  and {Berthold}, T., ``{J004457+4123 (Sharov 21): not a remarkable nova in M
  31 but a background quasar with a spectacular UV flare},'' {\em \aap}~{\bf
  512},  A1 (Mar. 2010).

\bibitem{2005A&A...429.1007S}
{Scholz}, A. and {Eisl{\"o}ffel}, J., ``{Rotation and variability of very low
  mass stars and brown dwarfs near {$\epsilon$} Ori},'' {\em \aap}~{\bf 429},
  1007--1023 (Jan. 2005).

\bibitem{PDC2015}
{Stecklum}, B., ``{Status of NEO confirmation observations at the Thüringer
  Landessternwarte (033)}.'' IAA Planetary Defense Conference, 2015,
  \url{http://iaaweb.org/iaa/Scientific%20Activity/conf/pdc2015/IAA-PDC-15-P-30.pdf}.
\newblock (Accessed: 24 May 2016).

\end{thebibliography}
\bibliographystyle{spiebib} 

\end{document}